# Superconducting hot-electron nanobolometer with microwave bias and readout

A. A. Kuzmin,[1,a)] M. Merker,[1] S. V. Shitov,[2,3,b)] ,N. N. Abramov,[2] A. B. Ermakov,[3] M. Arndt,[1] S. H. Wuensch,[1] K. S. Ilin,[1] A. V. Ustinov[2,4] and M. Siegel,[1]

[1] *Institut für Mikro- und Nanoelektronische Systeme, Karlsruhe Institut für Technologie KIT, 76187, Karlsruhe, Germany*

[2] *Laboratory for Superconducting Metamaterials, National University of Science and Technology MISIS, 119049, Moscow, Russia*

[3] *Kotel'nikov Institute of Radio Engineering and Electronics, 125009, Moscow, Russia*

[4] *Physikalisches Institut, Karlsruhe Institut für Technologie KIT, 76131 Karlsruhe, Germany*

We propose a new detection technique based on radio-frequency (RF) bias and readout of an antenna-coupled superconducting nanobolometer. This approach is suitable for Frequency-Division-Multiplexing (FDM) readout of large arrays using broadband low-noise RF amplifier. We call this new detector RFTES. This feasibility study was made on demonstrator devices which are made in all-Nb technology and operate at 4.2 K. The studied RFTES devices consist of an antenna-coupled superconducting nanobolometer made of ultrathin niobium films with transition temperature $T_c \approx$ 5.2 K. The 0.65-THz antenna and nanobolometer are embedded as a load into a GHz-range coplanar niobium resonator ($T_c \approx$ 8.9 K, $Q \approx$ 4000). To heat the superconducting Nb nanobolometer close to the $T_c$, the RF power at resonator frequency $f \approx$ 5.8 GHz is applied via a transmission line which is weakly coupled (-11 dB) to the loaded resonator. The THz-antenna of RFTES was placed in the focus of a sapphire immersion lens inside a He$^4$-cryostat equipped with an optical window and a semiconductor RF amplifier. We have demonstrated optical response of the RFTES to THz radiation. The demonstrator receiver system employing the RFTES device showed an optical Noise-Equivalent Power (NEP) ~$10^{-14}$ W/√Hz at 4.2 K.

## I. INTRODUCTION

Future space missions for THz astrophysics will have actively cooled primary mirrors that provide ultra-low background noise [1, 2.] These missions will require ultra-sensitive detectors with Noise Equivalent Power (NEP) ~$10^{-19} \div 10^{-21}$ W/√Hz for imaging spectrometers with large-format detector arrays (>10 kilopixel), which should be stable under cosmic ionizing radiation .

___________________________

a) Electronic mail: artem.kuzmin@kit.edu.

b) Electronic mail: sergey3e@gmail.com.

There are several competing concepts of ultrasensitive superconducting Terahertz direct detectors under consideration. These are membrane-based Transition-Edge Sensor (TES), the antenna-coupled Hot-Electron nanobolometers (nano-HEB) and Microwave Kinetic-Inductance Detectors (MKIDs). The fundamental noise limit for any bolometer is set by thermal fluctuations or phonon noise. The noise equivalent power (NEP$_{ph}$) defined by these fluctuations is:

$$\text{NEP}_{ph} = \sqrt{4k_B T^2 G} \ , \qquad (1)$$

where $T$ is the operation temperature and $G$ is the thermal conductance. At ultra-low temperatures, the number of photon and phonon cooling channels determines $G$. The benchmark for $G$ of a single-mode channel is the thermal conduction quanta[3, 4] $G_Q = \pi^2 k_B^2 T / 3h$ ($G_Q \approx 10^{-13}$ W/K at 100 mK). In principle it is possible, to get $G$ below $G_Q$ by introducing electromagnetic and acoustic mismatch in the cooling channels. In case of a membrane-based TES, this requires fabrication of narrow and very long dielectric supports for the membrane[5]. Therefore scaling up and dense packing of the pixels becomes a complicated task. In case of antenna-coupled nano-HEB made of disordered superconductor with Andreev mirrors, $G$ is mostly determined by electron-phonon interaction with characteristic time $\tau_{e\text{-}ph}$. The minimum NEP of nano-HEB is given by[6]:

$$\text{NEP}_{ph} = \sqrt{4\gamma k_B T^5 V / \tau_{e\text{-}ph}} \qquad (2)$$

where $V$ is the volume of the nano-HEB and $\gamma$ is the Sommerfeld constant. The electron-phonon time increases at lower temperature[7] as $\tau_{e\text{-}ph} \propto T^{-n}$, where $n \approx 2 \div 4$. Therefore, ultimate NEP$_{ph}$ values in the range of $10^{-20}$ W/$\sqrt{\text{Hz}}$ can be reached by decreasing the volume of the absorber down to nanometer scale and the temperatures down to milli-Kelvin range[8]. Additionally, a small volume of the sensor ensures a small influence of cosmic ionizing radiation[9]. The well-known issue of large arrays of any type of bolometer is the complexity of the readout system. TES and nano-HEB arrays rely on low-frequency SQUID-multiplexing that is a quite complicated and expensive solution[10].

On the other hand the pair-breaking MKID is free of the above-mentioned challenges of TES and HEB devices since it employ a coplanar waveguide (CPW) superconducting high-Q resonators coupled to a frequency-division multiplexing (FDM) readout transmission line. Therefore, MKID offers the capability for large scale multiplexing using a single broadband semiconductor low-noise amplifier[11] (LNA). The fundamental noise limit of MKID is set by the generation-recombination noise[12] (GR noise):

$$\text{NEP}_{gr} = \frac{2\Delta}{\eta_{opt}\eta_{pb}} \sqrt{\frac{n_{qp} V}{\tau_{qp}}} \propto \Delta \times n_{qp} \sqrt{V} \ , \qquad (3)$$

where $\Delta$ is the superconducting energy gap, $\eta_{opt}$, $\eta_{pb}$ are the optical-coupling and pair-breaking efficiencies, correspondingly, $V$ is the volume of the sensitive element, $n_{qp} \approx 2N_0\sqrt{2\pi k_B T \Delta_0}\exp(-\Delta_0/k_B T)$ is the quasiparticles density at $T \ll T_c$ ($N_0$ is the single-spin density of states at the Fermi energy), and $\tau_{qp} \propto \left(R_{rec} n_{qp}\right)^{-1}$ is the quasiparticle recombination time (here $R_{rec}$ is a material dependent recombination constant). At operation temperatures of a MKID, which is typically in the milli-Kelvin range, $\mathrm{NEP_{gr}}$ can reach extremely low levels since the number of thermally excited quasiparticles decreases exponentially with temperature. Aluminum-based MKID at 120 mK operation temperature have demonstrated already optical $\mathrm{NEP}\sim10^{-19}\,\mathrm{W}/\sqrt{\mathrm{Hz}}$ using amplitude readout[13]. Further improvement of NEP requires a lower readout power, $P_{readout}$, to reduce number of quasiparticles and/or a smaller volume of the sensitive element. Reduction of readout microwave power will require amplifiers with noise temperatures $T_n$ smaller than typical values of semiconductor LNA, otherwise NEP of MKID-based receiver will be limited by the amplifier noise[14, 15]:

$$\mathrm{NEP_{amp}} \approx \sqrt{\frac{k_B T_n}{2P_{readout}}} \times S^{-1}, \qquad (4)$$

where $S$ is the optical responsivity of MKID.

**Detector concept**

We have proposed to combine the advantages of nano-HEB and MKID[16, 17] by embedding an antenna-coupled superconducting nanobolometer into a quarter-wave ($\lambda/4$) high-Q superconducting resonator (Fig. 1 a). The resonator is weakly coupled to a transmission feed line at its shorted end. To match the impedance of the CPW resonator, the nanobolometer together with the antenna is placed close to the open end of the resonator. Operation bath temperature, $T_b$, for RFTES is below but near $T_c$ of the nanobolometer and well below $T_c$ of the resonator. The high Q-factor of the superconducting resonator ensures that nanobolometer is well isolated from external electrical noise and interference at all frequencies except close to resonance. Since operation conditions are analogous to operation conditions of the TES, we called this concept RFTES. If the nanobolometer is fabricated from a highly disordered thin-film superconductor, which has characteristic electron-electron interaction time $\tau_{e-e}$ much shorter than $\tau_{e-ph}$, a THz or RF signal with frequency $f \geq 1/\tau_{e-ph}$ will heat the quasiparticles[18] and consequently increase their number. In other words, only the electron subsystem of the nanobolometer serves as an absorber of the radiation with cooling time constant $\tau_{e-ph}$.

When the RF power is applied to terminal 1 of the RFTES (Fig. 1 a), it partially reflects and causes a dip in the transmittance to terminal 2, which is described by the scattering parameter $|S_{21}|^2(f)$ (Fig. 1 b). The nanobolometer absorbs another part of the power, $|S_{31}|^2$, which heats its electron gas. By increase of power at terminal 1, it is possible to heat electron gas of the nanobolometer close to $T_c$, where its surface resistance becomes strongly temperature dependent. The electromagnetic simulation at a resonance frequency of RFTES (shown in Fig. 1 c) demonstrate that, the higher surface resistance of the nanobolometer the higher the transmittance, $|S_{21}|^2_{min}$, of the feed line will be. Additional small heating of the nanobolometer by the THz radiation will thus proportionally increase the transmittance $|S_{21}|^2_{min}$ (dashed line on Fig. 1 (b)). The semiconductor broadband LNA connected to the terminal 2 of the transmission line then amplifies the output RF signal.

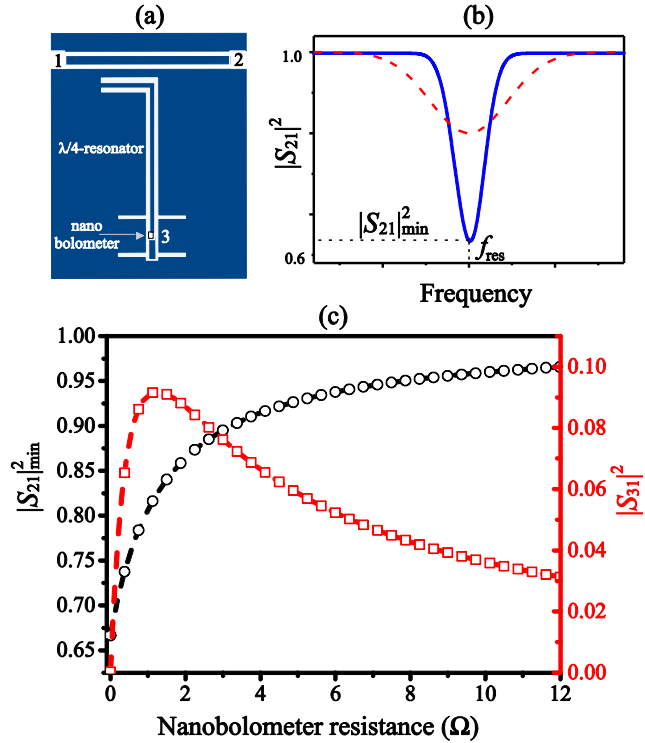

FIG. 1. a) Schematic 3-port (2-terminal) layout of the RFTES. The ports 1 and 2 are on the feed line. The port 3 represents the antenna-coupled nanobolometer; b) Schematic transmission spectrum $|S_{21}|^2(f)$ of RFTES at low (solid) and high (dashed) resistance of nanobolometer; c) The electromagnetic simulations of the 3-port model of RFTES at resonance frequency: dependence of transmittance $|S_{21}|^2_{min}$ (circles) and absorbed part of RF power $|S_{31}|^2$ (boxes) on resistance of nanobolometer.

The intrinsic responsivity of the device is the change of $|S_{21}|^2_{min}$ per unit of absorbed power in the nanobolometer $P_{bol}$ according to $S = d|S_{21}|^2_{min}/dP_{bol}$. The detailed analysis of the RFTES operation and its noise performance, based on the electromagnetic model, RF-superconducting and thermal properties of the nanobolometer and resonator will be published elsewhere.

## II. EXPERIMENTAL RESULTS

The presented feasibility study of the RFTES concept is made at liquid-helium temperatures using demonstrator devices based on Nb. The nanobolometer ($\sim 1.0 \times 0.46$ μm$^2$) with $T_c \approx 5.2$ K was made from a thin 15-nm thin-Nb film, deposited on sapphire substrate. The λ/4 resonator and the transmission line were made from 200-nm Nb film.

The thermal conductance $G$ of the nanobolometer was estimated leads using "isothermal" technique[19] from DC current-voltage characteristics of the witness bridge, measured at different bath temperatures, $T_b$. The witness bridge had the same size as nanobolometer and was fabricated on the same chip in the same patterning process. The value of $G \approx 1.25 \times 10^{-7}$ W/K was obtained on a sapphire at $T_b \sim 5$ K. The lower limit for NEP of the demonstrator RFTES was calculated from eq. (1) as $\mathrm{NEP}_{ph} \approx 10^{-14}$ W/$\sqrt{\mathrm{Hz}}$. To calculate contributions to NEP from other noise sources, the responsivity of the RFTES has to be estimated first.

### A. Microwave characterization.

The transmission spectrum $|S_{21}|^2(f)$ of the RFTES was measured using a vector network analyzer and He$^4$ dipstick. The resonance dip of the resonator was found around the designed frequency of 5.8 GHz (Fig. 2 (a)). However, $|S_{21}|^2_{min}$ was higher than obtained from simulations (Fig. 1 (b)). The resonance frequency and the Q-factor were dependent on bath temperature $T_b$ in accordance with change of surface impedance of thick Nb film at low input power. The loaded quality factor of the resonator $Q_L \approx 4 \times 10^3$ was achieved at 4.2 K. At a bath temperatures $T_b = 0.97 T_c$ and input power below critical value $P_{cr} \approx 20$ nW the transmittance $|S_{21}|^2$ is found to be independent on applied RF power. This corresponds to situation when both a heating effect is negligible and RF currents in the nanobolometer are below its critical current. However, when $P_{in} > P_{cr}$ the heating effect becomes significant. The measured transmittance increases with increase of input power, as depicted in Fig. 2 (a) and (b). In the framework of our RFTES concept, this behavior is due to increase of the surface resistance of the nanobolometer, which is caused by its uniform RF heating. When $P_{in}$ exceed saturation power $P_{sat} \approx 100$ nW (see Fig. 2 (b)) transmittance $|S_{21}|^2$ only slightly changing with increase of $P_{in}$. This corresponds to the normal state of the Nb nanobolometer, where its resistance weakly changes with temperature.

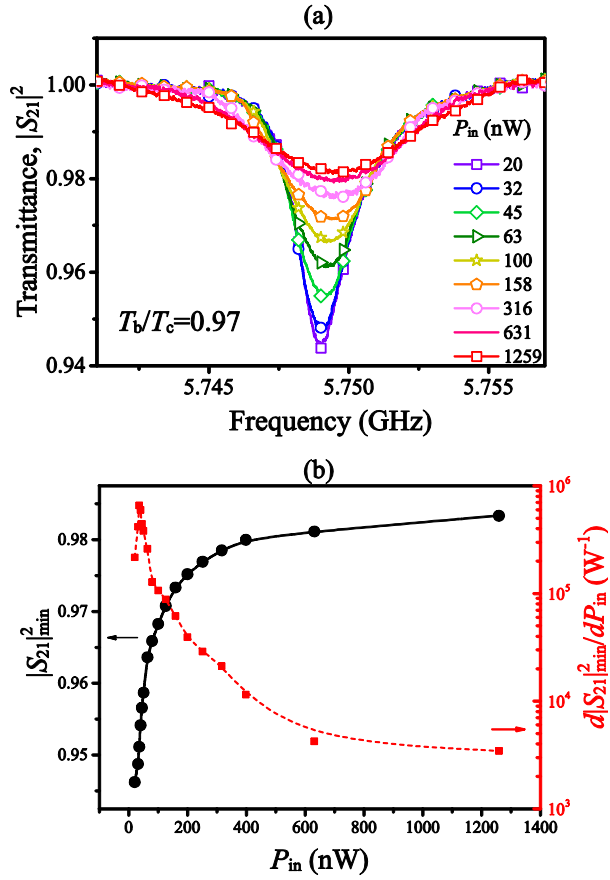

FIG. 2. a) Resonance curves of the RFTES for different input RF powers indicated in the inset and bath temperatures $T_b$=5.06 K ($T_b/T_c$=0.97); b) Dependence of the transmittance $|S_{21}|^2_{min}$ and its derivative $d|S_{21}|^2_{min}/dP_{in}$ on input RF power.

To estimate the intrinsic responsivity $d|S_{21}|^2_{min}/dP_{bol}$ from the measured dependence $d|S_{21}|^2_{min}/dP_{in}$ (Fig. 2 (b)) we can use coefficient $|S_{31}|^2 = P_{bol}/P_{in}$. It was roughly calculated for the point of saturation as $|S_{31}|^2_n \approx P_{bol}/P_{sat}$, when nanobolometer is almost in the normal state (electron temperature $T_e \approx T_c$) using the measured dependence $P_{sat}(T_b)$ (Fig. 3) and thermal conductance $G \approx 10^{-7}$ W/K. The estimated value of $|S_{31}|^2_n \approx 0.03$ for the bath temperatures $T_b < 0.92T_c$ (Fig. 3) is in good agreement with the value obtained from EM simulation for the resistance of the nanobolometer $R_{bol} \geq 10$ Ω (FIG. 1 (c)).

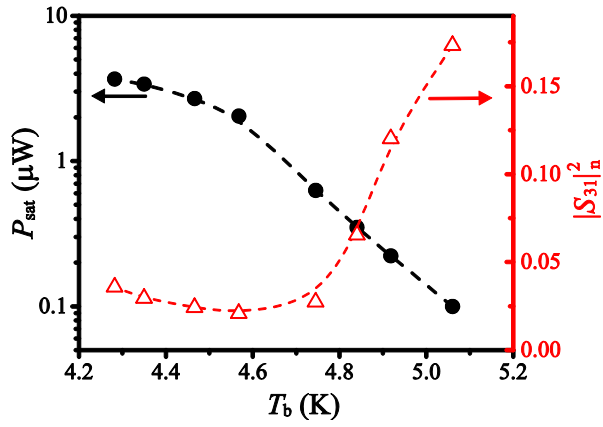

FIG.3. Temperature dependencies of the saturation $P_{sat}$ power (circles) and of coefficient $|S_{31}|_n^2$ (triangles) calculated from the temperature dependence of the saturation power.

However, for higher bath temperatures the $|S_{31}|_n^2$ increases. At the present time such behavior is not clear and has to be investigated further. The intrinsic responsivity of the RFTES is estimated to be $S = 10^6 \div 10^7$ W$^{-1}$. The contribution from the amplifier noise to the total NEP is $\text{NEP}_{amp} \approx 4 \times 10^{-15}$ W/$\sqrt{\text{Hz}}$ has been estimated from Eq. (4) for a noise temperature of the amplifier $T_n = 7$ K and $P_{reaout} = 100$ nW. The estimated contribution from GR noise of Nb resonator with the volume $V \approx 10^3$ µm$^3$ is $\text{NEP}_{GR} \approx 10^{-17}$ W/$\sqrt{\text{Hz}}$ Therefore, we can expect that the NEP of our all-Nb demonstrator RFTES is about $10^{-14}$ W/$\sqrt{\text{Hz}}$ and is dominated by thermal fluctuations in the nanobolometer.

**B. Optical response.**

To demonstrate the optical response of the RFTES device we used an experimental setup schematically shown in Fig. 4 (a). The RFTES is placed onto a sapphire lens mounted in a detector block which is connected to a cryogenic LNA (gain ~20 dB, $T_n$ ~7 K at 4-8 GHz) and installed inside a He$^4$ bath cryostat with a THz window. Blackbodies with known temperature were used as THz sources outside the cryostat. The radiation from the blackbody was filtered by a HDPE window and Zitex G106[20] foil at 4.2-K shield inside the cryostat. By design, the bandwidth of the double-slot antenna is about 100 GHz. This effectively restricts the spectrum of the detected signal. Only one spatial mode of the blackbody radiation is coupled because of the diffraction limited design of the extended hemispherical lens-antenna. The system response to radiation from blackbodies is shown in Fig. 4 (b). The maxima of the responses occur at slightly different bias powers. The sign of the shift for the optimal bias point depends on either the blackbody's temperature is higher or lower compared to room temperature.

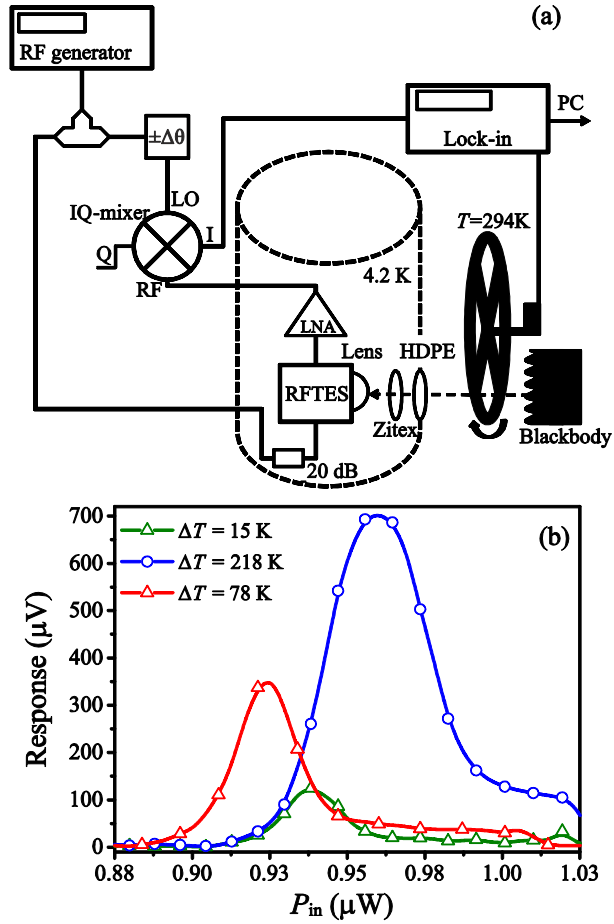

FIG. 4. a) Experimental setup based on IQ mixer for measurements of optical response of RFTES. b) Response of the RFTES to radiation from 3 different blackbodies i.e. human hand $T=310$ K, liquid nitrogen $T = 77$ K, boiling water $T = 373$ K in respect to room temperature $T=294$ K. The corresponding differences in temperature are indicated in the graph.

The accurate measurement of the optical NEP was made in another setup using a calibrated cryogenic blackbody radiator, which was placed near the lens of the detector block inside a pulsed-tube cooler. Details of this experiment will be discussed elsewhere. The incident THz radiation power from the cryogenic blackbody was calculated for the diffraction limited detector taking into account the antenna bandwidth. The optical NEP $\approx 3\times10^{-14}$ W/$\sqrt{\text{Hz}}$ of the demonstrator RFTES was obtained in this experiment, which is in good agreement with our estimation, discussed above.

### III. CONCLUSIONS

We tested a concept of a new THz detector with a microwave bias and readout - RFTES. The feasibility study with the demonstrator RFTES operating at 4.2 K successfully confirmed the proposed basic concept. The RFTES shows optical response to THz radiation with NEP $\approx 3\times10^{-14}$ W/$\sqrt{\text{Hz}}$ at $T = 4.2$ K. This value is in good agreement with estimations of the NEP of RFTES, which show dominant contribution of thermal fluctuation noise in the nanobolometer. A further reduction of $T_c$ and the volume of the nanobolometer should reduce NEP level defined by thermal fluctuations. To reach ultra-low NEP values $\sim 10^{-19} \div 10^{-20}$ W/$\sqrt{\text{Hz}}$ one could use highly disordered thin-film superconductors with $T_c \sim 100$ mK and reduce the volume of the nanobolometer down to $\sim 10^{-3}$ μm$^3$.

## ACKNOWLEDGMENTS

This work was supported in parts by grant 12-02-01352-a from Russian Foundation for Basic Research; Increase Competitiveness Program of NUST «MISiS» (№ K2-2014-025) and contract 11.G34.31.0062 from the Ministry of Education and Science of the Russian Federation; grants 05K13VK4, 13N12025 from German Federal Ministry of Education and Research (BMBF) and project № 284456 of the European Commission.